\documentclass[twocolumn]{aastex631}
\usepackage{amsmath}

\expandafter\ifx\csname natexlab\endcsname\relax\fi
\providecommand{\url}[1]{\href{#1}{#1}}
\providecommand{\dodoi}[1]{doi:~\href{http://doi.org/#1}{\nolinkurl{#1}}}
\providecommand{\doeprint}[1]{\href{http://ascl.net/#1}{\nolinkurl{http://ascl.net/#1}}}
\providecommand{\doarXiv}[1]{\href{https://arxiv.org/abs/#1}{\nolinkurl{https://arxiv.org/abs/#1}}}
\shorttitle{Malmquist-Like Bias in Caustic Areas and Time Delays}
\shortauthors{Baldwin \& Schechter}
\graphicspath{{./}{figures/}}
\begin{document}

\title{A Malmquist-like bias in the inferred areas of diamond caustics and
consequences for
inferred time delays of gravitationally lensed quasars}

\correspondingauthor{Paul L. Schechter}
	\email{schech@achernar.mit.edu}
	
	\author[0000-0002-4770-297X]{Derek M. Baldwin}
	\affiliation{MIT Department of Physics \\
		Cambridge, MA 02139, USA}
	
	\author[0000-0002-5665-4172]{Paul L. Schechter}
	\affiliation{MIT Department of Physics \\
		Cambridge, MA 02139, USA}
	\affiliation{MIT Kavli Institute for Astrophysics and Space Research \\
		Cambridge, MA 02139, USA}



\begin{abstract}
%\noindent
Quasars are quadruply lensed only when they lie within the diamond
caustic of a lensing galaxy.  This precondition produces a
Malmquist-like selection effect in observed populations of quadruply
lensed quasars, 
overestimating
the true caustic area.  The bias toward
high
values
of the
inferred logarithmic area, $\ln A_{inf}$, is proportional
to the square of the error in that area, $\sigma^2_{\ln{A}}$.  In
effect, Malmquist's correction compensates {\it post-hoc} for a
failure to incorporate a prior into parameter optimization.  Inferred
time delays are proportional to the square root of the inferred
caustic area of the lensing galaxy.  Model time delays are biased
long,
leading to
overestimats of the Hubble constant.  Crude estimates of
$\sigma_{\ln A}$ for a sample of 13 quadruple systems give a median
value of 0.16.

We identify a second effect, ``inferred magnification bias,'' resulting
from the combination of selection by apparent magnitude and errors in
model magnification.  It is strongly anti-correlated with caustic area
bias, and almost always leads to
underestimates
of the Hubble constant.
Malmquist's scheme can be adapted to priors on multiple parameters,
but for quad lenses, the negative covariances between caustic area and
absolute magnitude are poorly known.  Inferred
magnification bias may even cancel out caustic area bias, depending upon
(among other things) the slope of the number magnitude relation for
the sample.

Proper correction for these combined effects can, in principle, be
built into Bayesian modeling schemes as priors, eliminating the need
for Malmquist-style approximation, but is likely to be challenging
in practice.
\end{abstract}

\keywords{galaxies: quasars --- gravitational lensing: strong, Malmquist bias, time delay cosmography}

\section{Introduction}
%\begin{figure*}[htb!]
%\plotone{3astroids.PNG}
\begin{figure*}[t]
  \centering
  \includegraphics[width=1.0\textwidth,height=0.6\textwidth]{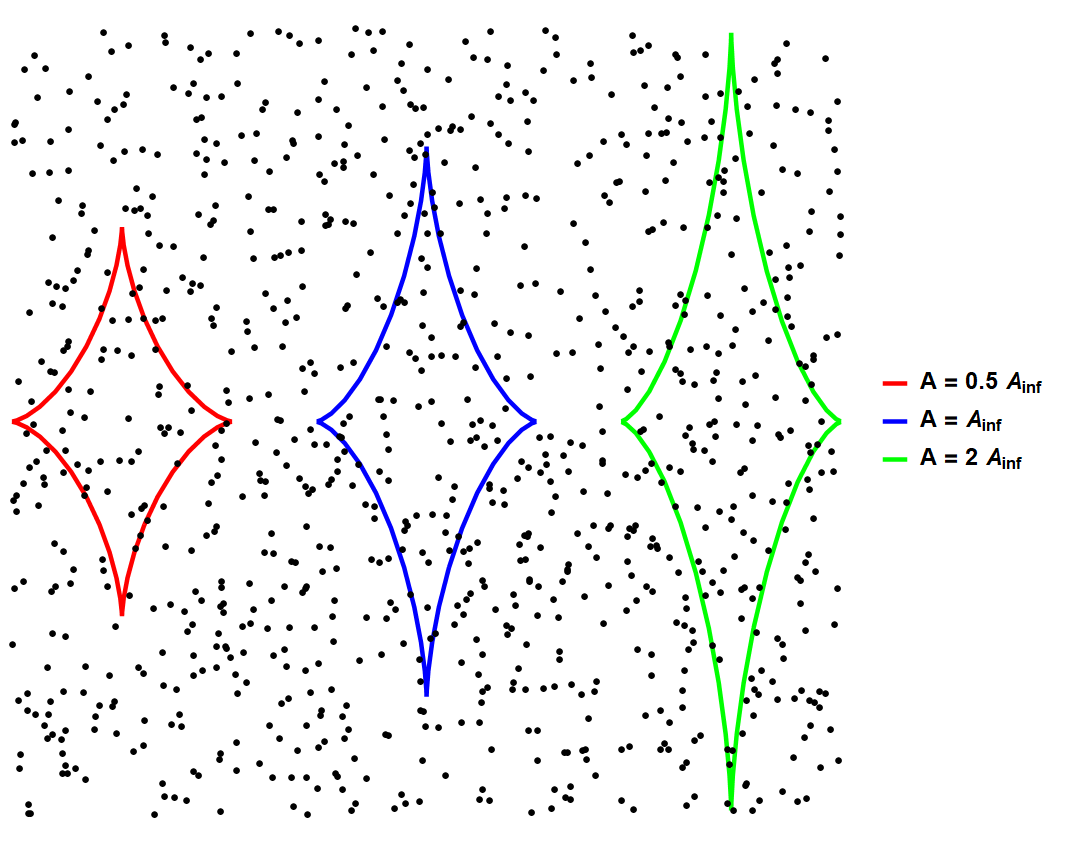}
\caption{Plots of diamond caustics with areas $\ln{A_{inf}} \pm
  \sigma_{\ln{A}}$, where $\sigma_{\ln{A}} = \ln{2}$, and randomly
  generated source quasar locations. Quads with higher caustic areas
  are over-represented in a sample because more source locations fall
  within the larger areas. The average caustic area for this sample of
  quads is $A_{peak}\approx 1.5 A_{inf}$.}
  \label{fig:figure1}
\end{figure*}
\indent The ``era of precision cosmology" (e.g. \citealt{kaplinghat})
was widely heralded by cosmologists (but not all; see
\citealt{bridle}) and the occasional astronomer.  Much of the
precision in ``precision cosmology" can be attributed to the high
degree of homogeneity and small amplitude of perturbations in the
early universe. But in today's universe the populations of
astronomical objects used for cosmological inference (e.g Cepheid
variable stars, type Ia supernovae, gravitationally lensed quasars)
exhibit wide ranges of properties that must be taken into full
account. These properties introduce uncertainties that become all the
more important with the ever-increasing precision of cosmological
measurements.

In this paper we discuss systematic biases in the measurements of
several quantities associated with gravitational lenses (specifically
caustic area, time delay, and magnification) that depend on these
uncertainties.  \indent We consider time delay measurements of
quadruply lensed quasars, widely regarded as a high precision method
for measurement of the Hubble constant, $H_0$
(\citealt{refsdal,treumarshall,wong}).  As is well-known, these
systems have a ``diamond" or ``astroidal\footnote{Note that these
  caustics form perfect astroids only in special cases.  The caustics
  considered here are almost, but not quite, stretched astroids.}"
caustic \citep{ohanian}, which gives rise to a ``quad" of images only
when the source quasar is located inside this caustic
\citep{finch}. We demonstrate that this condition leads to an analogue
of the well-known Malmquist effect \citep{Malmquist, binney}, in which
the true luminosities of stars in flux-limited samples are higher than
inferred
from their spectra
due to a selection bias in the sample. In Section 2 we show how a
similar selection effect occurs in the study of quadruply-lensed
quasars,
causing the true areas in ``quadruplicity'' selected samples
to be lower than inferred from modeling observations
of the lensed system.
We call this ``inferred caustic area bias.''
In Section \ref{sec:gedanken}
we describe a thought experiment showing that there is no
way to account for this bias without explicitly modeling it.
In Section \ref{sec:calculating} we derive formulae for  caustic areas
for several idealized models.  In Section \ref{sec:howbig} we use these
to estimate inferred caustic area bias in two samples.

While such an underestimate
in the inferred caustic areas of lensing galaxies might seem benign,
we argue in Section 6 that the area and associated relative time delay in
the system are strongly correlated, so that such an underestimate of
caustic area introduces an underestimate of the time delay for each
system. We claim that this in turn creates a finite
overestimage
of the Hubble constant that must be accounted for when making precise
measurements.  In Section 7 we show how this affects the Hubble constant.

Finally, in Section \ref{sec:magbias}, we summarize the results of
Appendix \ref{app:magmalm}, showing how random uncertainties in the
inferred magnification combine coherently with the uncertainties in
inferred caustic area (for the restricted case of perfect
anti-correlation) to alter the estimates obtained in the previous
sections.

All of our estimates are made using Malmquist's approximate, {\it
  post-hoc} approach to correcting the results of fitting schemes.
Malmquist's method grows increasingly unwieldy as one allows for
pairs, triplets and higher order multiplets of mutually correlated
parameters.  Many of these shortcomings might be avoided, at least in
principle, by incorporating the associated selection effects as
Bayesian priors within model fitting schemes.  But such Bayesian
schemes are often computationally expensive, and Malmquist's approach
permits a less expensive qualitative exploration of alternative schemes
for computing caustic area bias.

\section{The Malmquist-like effects for lens caustic areas}

\subsection{Classical Malmquist bias}

Malmquist's original paper \citep{Malmquist} concerned the inferred and true
intrinsic luminosities of stars included in a magnitude limited
sample.  The absolute magnitude of a star, $M_{inf}$, can be inferred
from examination of its spectrum.  But there is an error in that
inferred magnitude, the amplitude of which is taken to vary randomly
from one star to the next, distributed with a Gaussian of known width
$\sigma_{M}$ about some ''true'' absolute magnitude $M$.

While Malmquist allowed for non-uniformity, 
we assume for simplicity that the stars in the sample have constant spatial number
density.
We take the volume searched for a star with
absolute magnitude $M_{inf}$ to be $V_{inf}$.  Then the volume searched
for a brighter star with
absolute magnitude $M_{inf} - \sigma_M$
will be larger by a factor of $\exp(2.3026 \sigma_M)$
than $V_{inf}$, and stars with absolute magnitude
$M_{inf} - \sigma_M$ will be over-represented in the sample by that same
factor.
Averaging over the distribution of uncertainties, one has
\begin{equation}
  <M>_{sample} = M_{inf} - 1.38 \sigma^2_M \quad,
\end{equation}

The most important feature of this equation,
distinguishing  Malmquist's effect from
statistical errors, systematic instrumental errors, and systematic
errors in the calibration of a distance indicator, is that it varies
as the {\it square} of the error $\sigma^2_{M}$.  Secondarily, it
does not depend upon sample size.  The Gaussian errors can make the
true absolute magnitude brighter or fainter than $M_{inf}$.  By
contrast, Malmquist's effect always makes the expected sample average brighter
-- even for a sample of one.

\citet{binney} give a thorough treatment of the effect.
Our calculation of caustic area bias parallels their development
in somewhat condensed form.
\subsection{Caustic area bias}\label{subsec:areabias}
\indent In the same way that uncertainties in the derived luminosities
of stars cause the true luminosities of the stars in a
magnitude-limited sample to be brighter than the derived luminosities,
the uncertainty in the derived area of a lens caustic causes the true
caustic area to be larger than the derived caustic area.  This is due
to the fact that a quasar cannot be quadruply-lensed unless the source lies 
within the diamond caustic \citep{finch}. The systems we see as
``quads" are likely to have true astroidal caustic areas larger than
inferred because lensing galaxies with true astroidal caustics that
are 1 standard deviation larger than inferred are over-represented and
lensing galaxies with true astroidal caustics that are 1 standard
deviation smaller than inferred are under-represented. Since we select
only systems with their source quasar located inside the diamond
caustic, the probability that a system is included in a sample of
``quads" is proportional to the true area of its caustic. This is
illustrated in Figure 1.

\indent The derivation here of the size of the effect is similar to
the derivation of Malmquist's effect in \citet{binney}. We introduce a
Gaussian distribution of systems with true caustic areas $\ln A$ about an
inferred mean value $\ln A_{inf}$.

Then the contribution to the total
number of observed quadruply-lensed systems, $N$, with respect to area
is given by:
\begin{equation}\label{eqn:equation2}
\frac{dN}{d\ln{A}} = \frac{n}{\sqrt{2\pi\sigma_{\ln{A}}^2}}\left(\frac{A}{A_{inf}}\right)
\exp \left[\frac{-(\ln{A}-\ln{A_{inf}})^2}{2 \sigma_{\ln{A}}^2}\right]\,
\end{equation}
where $n$ is a number density (numbers per steradian) and
$\sigma_{\ln{A}}$ is the standard deviation in $\ln{A}$. The ratio
$A/A_{inf}$ gives the probability of observing a lens with true
caustic area $A$ relative to one with $A_{inf}$. Substituting
$\exp{\left(\ln{A}-\ln{A_{inf}}\right)}$ for $A/A_{inf}$
and letting ${\cal N} = n/\sqrt{2\pi\sigma_{\ln A}^2}$ we have
\begin{align}
  \frac{dN}{d \ln A} 
  & = {\cal N}
  \exp
  \left[
     \frac{-(\ln{A}-\ln{A_{inf}})^2}    
         {2 \sigma_{\ln{A}}^2}
  +(\ln{A}-\ln{A_{inf}})
  \right]
  \, \nonumber \\
&= {\cal N} \exp{\frac{-(\ln{A}-\ln{A_{inf}}-\sigma_{\ln{A}}^2)^2+\sigma_{\ln{A}}^4}{2 \sigma_{\ln{A}}^2}}\, \nonumber \\
&= {\cal N}  \exp{\frac{\sigma_{\ln{A}}^2}{2}} \exp{\frac{-(\ln{A}-\ln{A_{inf}}-\sigma_{\ln{A}}^2)^2}{2 \sigma_{\ln{A}}^2}} \nonumber
\end{align}
which is a new Gaussian distribution with a shifted peak at
\begin{equation}\label{eqn:areapeak}
\ln{A_{peak}} = \ln{A_{inf}}+\sigma_{\ln{A}}^2\,.
\end{equation} \label{eqn:defDELTAlnA}
We can then define the ``logarithmic area bias",
$\Delta_{\ln{A}}$, that reflects this systematic error:
\begin{equation}\label{eqn:causticBIAS}
    \Delta_{\ln{A}} \approx  \ln{A_{inf}} - \ln{A_{peak}} = -\sigma_{\ln{A}}^2 \,.
\end{equation}
where $A$ is the ``true'' caustic area that would be measured
if there were no error in the imperfect caustic area inferred
from the observations.
\indent The logarithmic mean of the caustic area inferred from
observations is larger than the 
is larger than inferred
by $\sigma_{\ln{A}}^2$ -- the inferred caustic areas are underestimated.
Similarly, Malmquist's effect causes inferred
intrinsic luminosities to be underestimated --
true absolute magnitudes are more negative
\begin{equation}\label{eqn:malmCORR}
    \Delta_{{M}}=M_{inf} - M_{peak} = 1.38 \sigma_{M}^2\,,
\end{equation}
where the factor of 1.38 is $\frac{3}{2}\ln{2.5}$. In both cases,
there is an additive bias in the logarithm of the relevant quantity --
absolute magnitude in the Malmquist case and logarithmic caustic area
for the quads -- that is proportional to the square in the uncertainty
in the logarithm of that quantity.

\subsection{``Correction'' or ``Bias''}\label{subsec:biasVScorr}

\citet{Malmquist} cast his scheme as a ``correction'' to fluxes
inferred from spectroscopy,
and equation (\ref{eqn:malmCORR}) gives that correction in magnitudes.
Fifty years later, \citet{Rubin1976} brought 
Malmquist's correction to bear on competing calculations of Hubble's constant,
which differed by a factor of two.  They used the words ``Malmquist bias''
to describe the consequence of not including the correction.

While Malmquist's ``correction'' scheme has fallen into disuse, the
term Malmquist ``bias'' has considerable currency, quite often in
discussions of the Hubble constant.  Had we adhered to Malmquist's
terminology, we would have called the quantity defined in equation
(\ref{eqn:malmCORR}), $\Delta_{M}$, the Malmquist {\it correction} and
the quantity defined in equation (\ref{eqn:causticBIAS}),
$\Delta_{\ln{A}}$, the logarithmic caustic area {\it correction}.
Instead, in what follows, we refer to these quantities as {\it
  biases}, a decision we came to regret only after the paper had been
refereed.  Readers are asked to remember that though we call these
quantities ``biases'', they are actually ``corrections,'' and that a
correction has a sign opposite that of a bias.

\section{A Bayesian interpretation of Malmquist's correction}

Malmquist's correction may be interpreted as an attempt to
account for
what Bayesians would call a prior on the absolute
magnitudes of stars selected by apparent magnitude.  The
prior has a simple mathematical form and, if one assumes
a Gaussian distribution of absolute magnitudes, the effect on
the mean absolute magnitude can be computed analytically.

Our calculation of caustic area bias is likewise a Bayesian
correction for a sample prior, again under the assumption
that the distribution of logarithmic caustic areas is Gaussian.  Multiplying
equation (\ref{eqn:equation2}) by $\ln A$ we
can likewise calculate the mean logarithmic caustic area by
integrating.

While that integral can done analytically, one could also do the
integral using Monte Carlo methods, if one preferred.  More generally,
if the distribution of logarithmic caustic areas can be
expressed analytically but is not Gaussian, one might integrate using
the trapezoidal rule or do Runge-Kutta integration.  And again, it
could also be done using Monte Carlo methods.

We have belabored what is obvious to most readers to emphasize
that one need not use Monte Carlo methods to account for
caustic area bias if one has a closed form solution for
the caustic in terms of the model parameters.

\section{Monte Carlo calculation of caustic area bias}

Figure \ref{fig:figure1} is, in effect, a Monte Carlo calculation of
caustic area bias for three different caustic areas.  We limited
the range of source positions sampled to the boundaries of our plot.

Lens modelers often use Markov chain Monte Carlo 
(henceforth MCMC) methods \citep{Sharma2017} that prefentially sample parameter values in some
neighborhood of the parameter values that are ultimately inferred.
They must then account for that non-uniform sampling.

If the correction for non-uniform sampling is perfect, the MCMC
modeling will have accounted for caustic area bias.  But corrections
for sampling bias are typically approximate, in which case there
will be finite residual caustic area bias.

\section{Joint calculation of caustic area bias and inferred magnification
bias}

Caustic area is strongly correlated with the magnifications inferred
for quadruply lensed images, which depend upon multiple
model parameters.

We discuss this at length in Section \ref{sec:magbias}, but
mention it here to indicate that Monte Carlo calculations
must take that strong coupling into account in correcting
for non-uniform sampling.  The coupling of caustic area
and inferred magnification will depend upon both source redshift
and lens redshift.  The priors for lens strength and intrinsic
source luminosity are poorly known.  Correction is challenging.
%We mention the issue at this point only to emphasize the challenges
%associated with correcting for caustic area bias using Monte Carlo
%methods.

\section{A {\it Gedanken} experiment}\label{sec:gedanken}

We circulated a post-submission, pre-publication version of this paper
and were surprised to find experts who thought that one or more of the
elaborate lens modeling programs presently used by lens modelers might
implicitly account for caustic area bias.

\subsection{Reductio ad absurdum}

We describe here a thought experiment that shows that if such an
experiment {\it could} detect caustic area bias, it could also break
the well known, mathematically exact ``mass sheet degeneracy''
%\citealt{falco,schneidersluse}.
\citep{falco,schneidersluse}.
Stated simply, one cannot tell, from lensed images alone, whether or
not a sheet of uniform mass surface density has been interposed
between a source and an observer, on either the near or far side.
This degeneracy can only be broken if one has additional information
about either the source (perhaps a type Ia supernova) or the lens
(e.g. by measuring the velocity dispersion of its stars).

We imagine a sample of gravitational lensed quasar systems each of
which has the same six identical model parameters -- an Einstein ring
diameter, an isothermal radial mass profile with an ellipticity and a
position angle, an external tidal shear with a different position
angle. and a seventh, variable parameter, the
true external convergence $\kappa_{ext}$.  The source
reshifts and lens redshifts are identical and each is known to infinite
precision.  We take the host galaxies to be identical and too small to be
resolved. Source positions are distributed randomly.

The observations of the quasars are subject to measurement errors,
which propagate to give uncertainties $\sigma_{\ln{A}}$
in the inferred logarithmic areas of the
diamond caustics.  As demonstrated in Section \ref{subsec:areabias}, the
mean inferred logarithmic caustic are will be overestimated.

Our Gedanken systems differ  only in the surface mass
density, parameterized by the external convergence
$\kappa_{ext}$,
which is proportional to the surface mass density of a
sheet in some fixed plane between the source and the observer.  The
logarithms of the convergences have a Gaussian distribution with
dispersion $\sigma^2_{\ln \kappa_{ext}}$ about a mean value.

The area of the diamond caustic is proportional to $1/(1 - \kappa_{ext})^2$.
If a lens modeling program could determine the area of the diamond caustic,
it would break the mass sheet degeneracy.
  
\subsection{Bayesian accounting for the mass sheet}\label{subsec:priorBETTER}

Our Malmquist-like correction for the mass sheets assumes the
Gaussianity of the distribution of logarithmic convergences.  There is
a straightforward (in principle) Bayesian alternative to Malmquist's
approach.  If one adopts a prior proportional to the caustic area $A$
in the modeling program, one can relax the assumption of Gaussianity
and choose a different scheme for weighting different possible values
for the convergence.

More generally, Malmquist's method of correction may be thought of as
using the derivative of the prior for some model parameter in the
vicinity of its most likely value and the derived variance for that
parameter to compute the effect of the prior.

\section{Calculating caustic areas -- idealized cases}\label{sec:calculating}

Though some lens modeling programs permit the plotting of the inferred
diamond caustic, none of those known to the authors when
this paper was first submitted produced areas for that
caustic or uncertainties in that area.  

This calculation is non-trivial.  The tangential caustic is diamondlike if the
lensing potential is symmetric about some axis, but is distorted if
that symmetry is broken, say by the presence of external shear that is not
aligned with the intrinsic flattening of the lens.

One can, however, derive approximate analytic expressions for several
idealized cases.  If the correctly calculated inferred caustic area is
not too different from these, one might as a first guess use the
scatter in these approximate caustic areas as a first guess for the
uncertainty in the correctly calculated caustic area.

Alternatively, one can find the critical curve numerically, project it
back onto the source plane and integrate either numerically or by
Monte Carlo methods.

\subsection{Singular isothermal potential in an external shear field}
     \citet{finch} give an expression for the caustic area of a singular isothermal potential in an external shear field, 
\begin{equation}
    \psi(r,\theta) = br + \frac{\gamma r^2}{2}\cos{2\theta}\,,
\end{equation}
and find
\begin{equation}
    A = \frac{3\pi}{2} b^2 \frac{\gamma^2}{1-\gamma^2}\,.
\end{equation}
This is one of the few lensing potentials for which an exact analytic solution is possible, and is 
a limiting case of the next potential we consider.
\subsection{Singular isothermal elliptical potential with parallel external shear}\label{subsec:SIEPXS}
\citet{luhtaru} consider the SIEP + XS$_{||}$ model,
\begin{equation}
    \psi(r,\theta) = b\sqrt{q_{pot}x^2 + \frac{y^2}{q_{pot}}} - \frac{\gamma}{2} (x^2-y^2)\,.
\end{equation}
They arrive at the approximate expression
\begin{equation}
    A \approx \frac{3\pi b^2}{2}(\gamma + 2\eta)^2\,,
\end{equation}
where $\eta$ is the semi-ellipticity, $q_{pot} = (1-\eta)/(1+\eta)$. They rewrite this in terms of an ``effective quadrupole,"
\begin{equation}
    \Gamma_{eff} = \frac{\gamma + \eta}{1 + \gamma \eta }\,,
\end{equation}
which is tightly constrained by the elongation of the image configuration. The parameter $b$, roughly the Einstein radius, $\theta_E$, is also well constrained.  But while $\Gamma_{eff}$ is well constrained, the ratio of the semi-ellipticity
to the shear is usually much more poorly constrained.

Substituting for the semi-ellipticity, we get
\begin{equation}
A \approx \frac{3\pi \theta_E^2}{2}\left(2\Gamma_{eff} - \gamma \right)^2\,.
\end{equation}
\indent As the shear, $\gamma$, is more poorly constrained than the
other parameters, it is the principal source of uncertainty in the
caustic area.  Taking the other parameters to be constant, an error
in the shear, $\delta \gamma$, produces an error in the inferred
logarithmic area, $\delta \ln A_{inf}$. We
have
\begin{equation}\label{eqn:dlnA}
   \delta \ln A_{inf} = \frac{2 \delta \gamma}{(2\Gamma_{eff} - \gamma)} \quad .
\end{equation}  

\subsection{Power-law + external shear potential}\label{subsec:PLXS}
Another manageable idealization is
the power-law + external shear potential,
\begin{equation}\label{eqn:PLXSpot}
  \psi(r,\theta) = b r^{\alpha} + \frac{\gamma r^2}{2}\cos{2\theta}\,,
\end{equation}
which has Einstein radius $\theta_E = (b\alpha)^{1/(2-\alpha)}$.
In Appendix \ref{app:powerlaw} we calculate the approximate area of the diamond caustic for this case,

\begin{equation}\label{eqn:PLarea}
    A \approx \frac{3\pi}{2}\theta_E^2 \gamma^2 \left(\frac{1}{1-\gamma^2}\right)^{\frac{1}{2-\alpha}}\,.
\end{equation}
\indent As noted previously, the Einstein radius is very well
determined for any given system and so is relatively constant even as
the slope is allowed to vary.  However one expects strong covariance
between the slope and external shear.

%\section{How big is the caustic area bias?}\label{sec:howbig}
\section{Practical uncertainties in calculated caustic areas}\label{sec:howbig}

We use two alternative schemes for modeling the typical
uncertainty in the logarthmic area from equation (\ref{eqn:dlnA}).
The first of these was developed by   
\citet{luhtaru}, who modeled the image and galaxy positions of 39
quasars lensed by relatively isolated galaxies with an isothermal
elliptical potential with external shear parallel to the ellipticity
(henceforth ${\rm SIEP+XS_{\parallel}}$).  They report shear,
semi-ellipticity, effective shear $\Gamma_{eff}$ and approximate
errors in the shear.  Using equation (\ref{eqn:dlnA}) we calculate
uncertainties in logarithmic caustic area, $\delta \ln A_{inf}$ and
find a median value of 0.321.

Alternatively, we use the results of 
\cite{Shajib}, who automated their modeling to give hands-off models for 13 quadruply
lensed quasars,
reporting model parameters and uncertainties
for shears and mass axis-ratios, each with a distinct
orientation.  As equation (\ref{eqn:dlnA}) was
derived for the case of shear parallel to ellipticity, we add the
shear and semi-ellipticity as spinors \citep{luhtaru} and consider
only their projected components.  Covariances were not reported,
so we take the uncertainties in the shear and semi-ellipticity to
be uncorrelated.  
Incorporating these into equation (\ref{eqn:dlnA}),
we get a median uncertainty in the logarithmic caustic area of
$\delta \ln A_{inf}$ = 0.16.

As might be expected, the uncertainties from this alternative model
(henceforth referred to as Bayesian) are somewhat smaller than
for the ${\rm SIEP+XS_{\parallel}}$ model.
The Bayesian model
has more free parameters than the latter, allowing for better
fits to the image position data.  
We note, however, that equation (\ref{eqn:dlnA}) was derived under
the assumption that projected shear and projected
semi-ellipticity are perfectly anti-correlated.  Proper
accounting of actual covariances would have different
effects on different systems.

A second reason to expect the Bayesian model to yield smaller
uncertainties is that it uses the extended structure of the quasar
hosts to constrain the model.
  
The approximations we used to estimate the uncertainties
in inferred cautic areas were crude.  To do better,
one would need to find the
critical curve, project it back onto the source plane and integrate
numerically in the course of carrying out the Bayesian calculation, so
as to account for of the covariance of shear and central concentration
discussed in Section \ref{subsec:PLXS}.

Summarizing our alternative schemes, we get uncertainties in the
inferred logarithmic caustic area of
\begin{equation}\label{alternateUNCS}
  \delta {\ln A_{inf}}  =
  \left
  \lbrace
  \begin{matrix}
    ~0.32~(\rm SIEP+XS_{\parallel}) \hskip 8pt {\rm or} ~ \cr
    ~0.16~(\rm Bayesian) \hskip 28pt . ~~  \cr
  \end{matrix}
  \right .
\end {equation}
\smallskip
\section{Implications of caustic area bias for time delays and the Hubble constant}\label{sec:implications}

\begin{figure*}[htb!]
\plotone{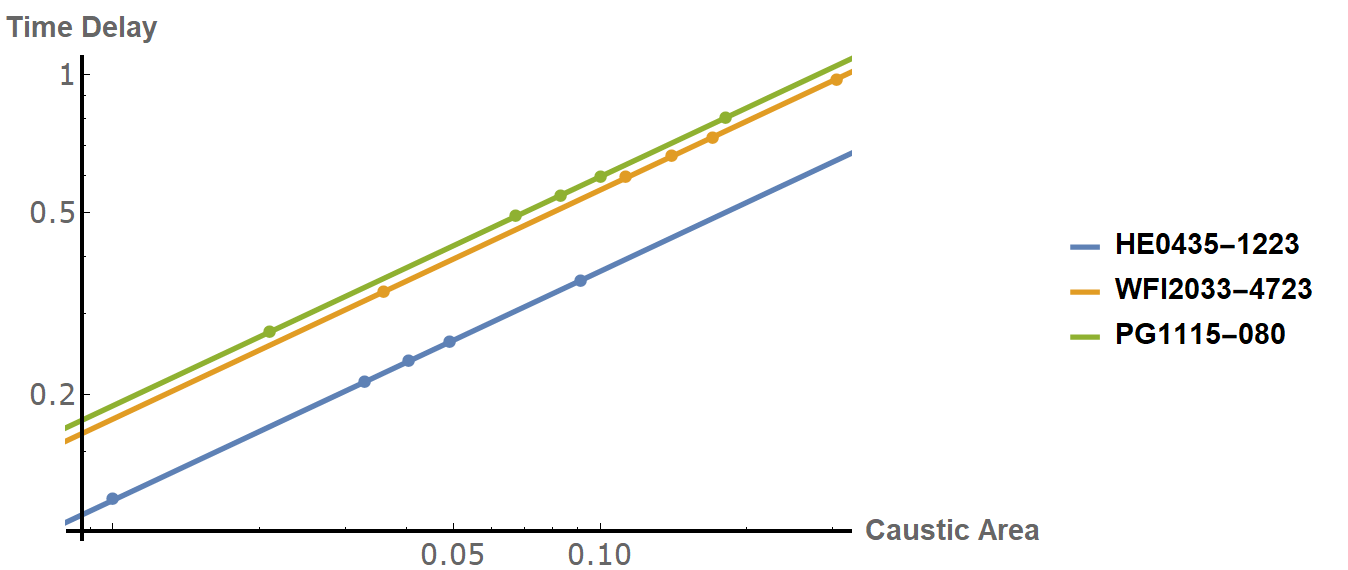}
\caption{Log-Log Plots for Caustic Area ($A$) vs. Time Delay ($\tau$) with lines of best fit. Each line of best fit has slope $\frac{1}{2}\,.$}
\end{figure*}

\subsection{Varying power-law slope}\label{subsec:powerlaw}
\indent We examine 3 different systems that figure prominently in
recent TDCOSMO  results \citep{Birrer}: HE0435--1223, WFI2033--4723, and
PG1115+080, and examine how the variation in the area of the diamond
caustic influences time delay.  We adopt the power-law + external shear
lensing potential of equation (\ref{eqn:PLXSpot}), as implemented in
Keeton's {\tt lensmodel} program \citep{lensmodel}.

\indent While {\tt lensmodel} permits the specification
of a fiducial Hubble constant and redshifts for the lens and source,
we use the program in its dimensionless mode.

{\parskip=0pt 
\begin{enumerate}
\item For each of the 3 systems, we find the lens model that best fits
  the observations (image/galaxy position, image fluxes, etc.), all
  with a power-law slope ($\alpha$) value of 1.
\item We then create fictitious ``perfect twin systems" with image
  positions, fluxes, etc. altered to exactly match the simulated
  values from {\tt lensmodel}. This removes uncertainties due to
  image/galaxy positions so as to isolate the effect of changes
  in  $\alpha$ on caustic area.
\item We next construct new models for each system with set $\alpha$
  values of 0.5, 0.9, 1.0, 1.1, and 1.5. Each new model is generated with
  the ``perfect twin" data values for the positions, fluxes, etc. This
  creates different-sized diamond caustics with areas given by
  equation (\ref{eqn:PLarea}).
\item Finally we find the time delays for each of these modeled
  systems. The caustic areas are calculated numerically from the
  curves generated by {\tt lensmodel}.
\end{enumerate}
}
The results for each of the 3 systems are contained below in Table 1 and Figure 2.
\startlongtable
\begin{deluxetable}{clllllr}
\tablenum{1}
\tablecaption{Areas and Time Delays for HE0435--1223, WFI2033--4723, and PG1115+080}
\tablewidth{0pt}
\tablehead{\colhead{} & \multicolumn{2}{c}{HE0435--1223} & \multicolumn{2}{c}{WFI2033--4723} & \multicolumn{2}{c}{PG1115+080}\\
\colhead{$\alpha$} & \colhead{$A$} & \colhead{$\tau$} & \colhead{$A$} &
\colhead{$\tau$} & \colhead{$A$} & \colhead{$\tau$} \\
\colhead{} & \colhead{$(\prime\prime)^2$} & \colhead{(Time)} & \colhead{$(\prime\prime)^2$} &
\colhead{(Time)} & \colhead{$(\prime\prime)^2$} & \colhead{(Time)}
}
%\decimalcolnumbers
\startdata
 0.5 & 0.0913 & 0.3544 & 0.305 & 0.9740 & 0.181 & 0.8048\\ 
 0.9 &  0.0491 & 0.2605 & 0.170 & 0.7278 & 0.100 & 0.5975\\
 1.0 & 0.0405 & 0.2370 & 0.140 & 0.6640 & 0.083 & 0.5444\\
 1.1 & 0.0328 & 0.2134 &  0.113 & 0.5994 & 0.067 & 0.4909\\
 1.5& 0.0100 & 0.1187 & 0.036 & 0.3354 & 0.021 & 0.2739\\
\enddata
\end{deluxetable}

We see in Figure 2 that all three cases have a straight line log-log
plot with slope $0.5 \pm 0.01$.  Without further proof, we shall assume
that the time delay follows the relationship $\tau \propto \sqrt{A}$.
If, as we have argued, the inferred caustic area is overestimated,
the inferred time delay is also overestimated.

\subsection{The effects of external convergences on time delays}

The diamond caustic is larger when there is a positive external
convergence than it would otherwise be.
But the value of such a convergence is completely indeterminate,
a consequence of the mass sheet degeneracy \citep{falco,schneidersluse},
unless non-lensing observations (i.e. counts of nearby
galaxies or measurements of the stellar velocity dispersion
in the lens) are brought into the modeling.
The position of the
source {\it relative} to the diamond caustic remains unchanged
by an external convergence.
\citep{FalorSchechter}.  The path lengths inferred from the lensing observations
are therefore all too short by
a factor $(1-\kappa_{ext})$, and the inferred time delays are
likewise too long by the same factor.

As inferred lengths scale as $(1 - \kappa_{ext})$, the inferred
area of the diamond caustic is too small by the factor
$(1 - \kappa_{ext})^2$.
The inferred time delay scales as the square root of the inferred
area, as was found experimentally for the case of the power-law potential
with external shear in Section \ref{subsec:powerlaw}. \\

\section{The effect of caustic area bias on the Hubble constant}

As discussed in Section \ref{sec:howbig},
the results of \citet{luhtaru} give uncertainties in the logarithmic
caustic area $\delta \ln A_{inf} \approx 0.32$, while the results
of \citet{Shajib} 
give uncertainties $\delta \ln A_{inf}\approx 0.16.$
From equation (\ref{eqn:causticBIAS})
we have
\begin{equation}\label{biasEST}
 \Delta \ln A  \approx   \frac{\ln A_{inf}}{\ln A} \approx
  \left
  \lbrace
  \begin{matrix}
    ~-0.10~~(\rm SIEP+XS_{\parallel}) \hskip 8pt {\rm or} ~ \cr
    ~-0.025~(\rm Bayesian) \hskip 28pt . ~~  \cr
  \end{matrix}
  \right .
\end {equation}
\smallskip
Note that $\Delta {\ln A}$, as defined in equation (\ref{eqn:causticBIAS})
and discussed in Section \ref{subsec:biasVScorr}, 
is the {\it correction}
to the inferred logarithmic caustic area that Malmquist would have
computed.  It is negative.

We argued in Section
$\ref{subsec:powerlaw}$
that if
the inferred caustic area is overestimated,
the time delay inferred from a perfectly measured
time delay, $\tau_{obs}$ is also overestimated, with
\begin{equation}
  \frac{\ln \tau_{inf}}{\ln \tau_{obs}}
  =  \frac{1}{2} \frac{\ln A_{inf}}{\ln A} \quad.    
\end{equation}
This leads to a positive bias in the inferred value of
the Hubble constant.  We have
\begin{equation}
  \frac{\ln H_0^{inf}}{\ln H_0^{true}} \approx 
  -\frac{1}{2}\frac{\ln A_{inf}}{\ln A_{true}} \approx
  \left
  \lbrace
  \begin{matrix}
    ~0.05~~(\rm SIEP+XS_{\parallel}) \hskip 8pt {\rm or} ~ \cr
    ~0.012~(\rm Bayesian) \hskip 28pt . ~~  \cr
  \end{matrix}
  \right .
\end{equation}
Hence the bias in the logarithm of the Hubble constant
has the same amplitude but the opposite sign as the bias in the time
delay.
%\indent
This is small compared to the random uncertainties in the
reported measurements of $H_0$, but it is systematic, and its sign is
always positive.  At such time as modelers estimate the caustic area
in the course of their modeling, we will have a better estimate of its
effect.  And as noted in Section \ref{sec:gedanken}, one can then
put a prior on caustic area that obviates the need for Malmquist's
correction.

In Section \ref{sec:howbig} we estimated the typical uncertainty in the logarithmic caustic area
for the models
of \citet{luhtaru} and \citet{Shajib}, respectively.  Equation (\ref{eqn:dlnA})
, 
used in
producing these estimates,
falls short for all of the systems in both papers, in some cases badly.
One cannot properly calculate the uncertainty in the caustic area
without properly calculating the caustic area itself within the modeling program.  The difference between the ${\rm SIEP+XS}_{\parallel}$ estimate and
the Bayesian estimate gives some sense of the residual systematic
error in the derived Hubble constant.

\section{The dependence of inferred caustic area bias on the anti-correlation
  of area and magnification}\label{sec:magbias}

In the preceding sections, we argued that when modeling a sample of
quadruply lensed quasar systems, the average true area of their
diamond caustics will be systematically larger than their average
inferred area, by an amount approximately proportional to the {\it
  square} of the typical individual error.  This systematic error is
the result of individual systems with positive errors ($\ln A -
\ln A_{inf}$) being quadruply lensed more often than those with
negative errors.

There is a {\it class} of systematic errors of this nature, the
archetype of which was identified by \citet{Malmquist}, who considered
the inferred absolute magnitudes for stars in a magnitude limited
sample.

We consider here another such systematic error associated with the
inferred magnifications of models for lensed quasars.  It is the
result of systems with positive magnification errors ($\mu -
\mu_{inf}$) penetrating further into the rising number-magnitude
relation for quasars than those those with negative errors.

The immediate temptation is to use Malmquist's method to make a crude
estimate of an ``inferred magnification bias''.\footnote{This is not
  to be confused with ``magnification bias", the term introduced by
  \citet{turner} to explain why the ratio of lensed quasars to
  unlensed quasars is higher at brighter apparent magnitudes.}  But
magnification and caustic area are in general strongly
anti-correlated.  They are, for example, inversely proportional to
each other as one varies the convergence of a mass sheet.  Proper
treatment requires knowledge of the covariance of caustic area and
magnification, and most likely other model parameters as well.

Malmquist's approach is poorly suited to this ``correlated
area-magnification bias''.  It might be more effectively addressed by the
inclusion of a Bayesian prior in the modeling program, as suggested in
Section \ref{subsec:priorBETTER}.

Notwithstanding its shortcomings, Malmquist's method yields an
important conclusion: that the bias due to selection by apparent magnitude
tends to mitigate the effects of selection by caustic area.

In Appendix \ref{app:magmalm} we derive expressions for
the systematic errors in the inferred magnification and
the inferred caustic area, under the following assumptions:
\begin{itemize}
\item Caustic area and magnification are perfectly anti-correlated, with $A
  \propto \mu^{-1/\alpha}$ and $\alpha > 0$, 
\item the redshifts of the lens and source are known,
\item the number magnitude relation in the vicinity
  of $m_{inf}$, the apparent magnitude associated the
  inferred magnifcation $\mu_{inf}$ is given by
  $dN/dm \propto 10^{\beta(m - m_{inf})}$ (which  increases
  at fainter apparent magnitudes for $\beta > 0$) and
\item the  errors in $\mu_{inf}$ are Gaussian.
\end{itemize}  
For the caustic area, we find
\begin{equation}
\ln A_{peak} = \ln A_{inf} - \sigma^2_{\ln A} +2.5 {\beta}{\alpha}\sigma^2_{\ln A}\quad , 
\end{equation}
which agrees with equation (\ref{eqn:areapeak}) if $\beta =0 $,
corresponding to a flat number-magnitude relation.

For a rising number-magnitude relation, $\beta > 0$,
ignoring the use of apparent magnitude to select a sample of
quads causes the inferred caustic area bias (as computed in the
previous sections) to be {\it overestimated}.

For the purpose of illustration (but not imitation) suppose $\beta =
0.3$ as might be appropriate for recently discovered quadruply lensed
quasars.  Suppose further that $\alpha = 1$, as would be appropriate
if the errors in inferred magnification were due to uncertainty in the
intervening mass sheet.  The bias in the inferred logarithmic
caustic area is then smaller by a factor of four than it would have
been had magnification not been taken into account.

The slope $\beta$ characterizing the number-magnitude relation is
unlikely to be the same for different classes of lensed sources.  
For example, magnitude selected samples of quasars and supernovae are
likely to have different systematic errors.

\section{Conclusion}

We have shown that, on average, the galaxies lensing quads will have
inferred caustic areas biased low with respect to the true caustic
areas.  This results from the proportionality of cross section for
quadruplicity to caustic area.  We presented a {\it Gedanken} experiment to
establish that the effect is inescapable.  We have used Malmquist's
approximate method to estimate the amplitude of the effect for two
sets of published model results.

There is a straightforward relationship between the caustic area of a
given system and its associated time delay, $\tau\propto
\sqrt{A}$.  A systematic bias in the inferred caustic areas produces
a systematic bias in the inferred time delay.  As the Hubble constant
is inversely proportional to the inferred time delay, this bias low in
area introduces a systematic underestimate of the inferred
Hubble constant.

This inferred caustic area bias is mitigated by a second
Malmquist-like bias due to errors in inferred model magnifications and
the selection of systems by apparent magnitude.  It appears to act
oppositely to  the caustic area bias, combining to produce a smaller net effect
because caustic area and magnification are anti-correlated.  Malmquist's
approximate {\it post-hoc} correction scheme is ill-suited to deal with
correlated parameters.  One can, for sufficiently restrictive cases,
calculate approximate corrections for the two selection effects from
reported uncertainties in (and reported covariances for) the model
parameters, but these are of limited practical use.

As an alternative to Malmquist's approach, one might in principle
incorporate Bayesian priors for caustic area and magnification
directly into model fitting programs, if the added computation does
not prove to be prohibitive.

\medskip
\centerline{ACKNOWLEDGEMENTS}
\medskip
The authors thank an anonymous referee for noting that Bayesian
Monte Carlo methods have the effect of correcting,
albeit incompletely,
for caustic area bias.  The authors thank Richard Luhtaru and
Drs.\ Philip Marshall, Sherry Suyu and Thomas Collett for comments on
the manuscript.  DB thanks the MIT UROP program for its support.

\appendix

\section{Approximate caustic area for a power law + external shear potential}\label{app:powerlaw}

Here we follow closely the development in \citet{finch} to derive an approximate analytic expression for astroidal caustic areas. We begin with the lensing potential: 
\begin{equation}\psi(r,\theta) = b r^{\alpha} + \frac{\gamma r^2}{2}\cos{2\theta}\,.\end{equation}
In Cartesian form:
\begin{align}
    \psi(x,y) &= b \left(x^2+y^2\right)^\frac{\alpha}{2}+\frac{\gamma \left(x^2+y^2\right)}{2}\cos{2\arctan{\frac{y}{x}}} \nonumber \\
    &= b\left(x^2+y^2\right)^\frac{\alpha}{2}+\frac{\gamma}{2}\left(x^2-y^2\right)\,.
\end{align}
The astroidal caustic is the locus of points for which the determinant
of the inverse magnification matrix,
\begin{equation}\mu^{-1} = \left(1-\frac{\partial ^2 \psi}{\partial x^2}\right) \left(1-\frac{\partial^2\psi}{\partial y^2} \right)-\left(\frac{\partial^2 \psi}{\partial x\partial y}\right)\end{equation}
disappears. \\
\indent As for the shape of the caustic, according to \citet{anevans}, the ``series [for the area] are also Taylor-series expansions with respect to $\gamma$ at $\gamma = 0$. If we truncate the expansion of the
caustic after [the first] term, the expression reduces to the equation of
the tetra-cuspi-hypocycloid, or the astroid." We make the simplifying assumption that $\gamma << 1$, so as to simplify the expression to that of approximately an astroid. The area of an astroid is given by
\begin{equation}A = \frac{3\pi}{8}x_c y_c\,,\end{equation}
where $x_c$ and $y_c$ are the $x$ and $y$ intercepts of the caustic (Finch et al. 2002). In what follows we consider only the intercepts of the caustic.
Ignoring the cross term, which goes to zero on the axes,  we are left with:
\begin{align}
\mu^{-1} &= \left\{1-b\left[x_c^2+y_c^2\right]^{\frac{\alpha-4}{2}}\left[\alpha y_c^2+\alpha(\alpha-1)x_c^2\right] - \gamma \right\} \nonumber \\
     & \times \left\{1-b\left[x_c^2+y_c^2\right]^{\frac{\alpha-4}{2}}\left[\alpha x_c^2+\alpha(\alpha-1)y_c^2\right] + \gamma \right\} \nonumber \\
\end{align}

For a source at a cusp, the magnification at the corresponding point
in the image plane is infinite in the direction transverse to that
point's displacement from the center of the lens.  For a cusp on
the $x$-axis, the magnification is infinite in the $y$-direction.
Hence the $1 - \partial^2 \psi/\partial y^2$ factors
in equations (A3) and (A5) must be zero, giving
\begin{align}
   0 = 1-b\alpha x_c^{\alpha-2}+\gamma\,,
\end{align} and
\begin{equation}
  x_c =\pm b^{\frac{1}{2-\alpha}} \left(\frac{\alpha}{1+\gamma}\right)^{\frac{1}{2-\alpha}}\,.
\end{equation}
A parallel calculation for the $y$-intercept gives us
\begin{equation}
    y_c = \pm b^{\frac{1}{2-\alpha}} \left(\frac{\alpha}{1-\gamma}\right)^{\frac{1}{2-\alpha}}\,.
\end{equation}
\indent Remapping back to the source plane via the lens equation, 
\begin{equation}\vec{r}-\vec{r_{s}} = \nabla\psi(\vec{r})\,,\end{equation}
we get 
\begin{align}
    x_a &= x_c \left(1-b\alpha x_c^{\alpha-2}-\gamma\right)\\
    &= -2\gamma x_c\\
    &= \pm 2\gamma b^{\frac{1}{2-\alpha}} \left(\frac{\alpha}{1+\gamma}\right)^{\frac{1}{2-\alpha}}\,.
\end{align}
The same process for the y-intercept yields 
\begin{align}
    y_c = \pm 2\gamma b^{\frac{1}{2-\alpha}} \left(\frac{\alpha}{1-\gamma}\right)^{\frac{1}{2-\alpha}}\,.
\end{align}
\indent By the assumption $\gamma<<1$, we know the area of the caustic is proportional to the product of these two intercepts. The constant of proportionality for an astroid shape is $3\pi/8$. Thus we arrive at the final expression
\begin{equation}
    A = \frac{3\pi}{2}\gamma^2 b^{\frac{2}{2-\alpha}} \left(\frac{\alpha^2}{1-\gamma^2}\right)^{\frac{1}{2-\alpha}}\,.
\end{equation}

\section{Systematic effects due to errors in inferred magnification}\label{app:magmalm}

As outlined in Section \ref{sec:magbias}, the application of Malmquist's method
to approximate the bias in inferred magnification and the resulting inferred
time delay is more complicated than for inferred caustic area bias.

We make the following assumptions:
\begin{itemize}
\item Caustic area and magnification are perfectly anti-correlated
 with $A \propto \mu^{-1/\alpha}$, $\alpha >0 $,
\item The redshifts of the lens and source are known,
\item The number magnitude relation in the vicinity 
  of $m_{inf}$, the apparent magnitude associated the
  inferred magnification $\mu_{inf}$, is given by
  $dN/dm \propto 10^\beta(m - m_{inf})$ with the expectation that $\beta > 0$ and
\item The errors in $\mu_{inf}$ are Gaussian. 
\end{itemize}  

We briefly suppress the additional complication of caustic area bias.  We
assume that the sample is comprised of lensed quasars with the same
observed apparent magnitude $m_{obs}$ over some small range $\pm {\cal
  D}m_{obs}$ and that we have selected a subsample of these that all
have the same inferred logarithmic magnification ${\ln\mu_{inf}}$ over
some small range $\pm {\cal D} \ln\mu_{inf}$.  This is the analogue,
in Malmquist's treatment, of choosing all the stars with the same
absolute magnitude, as inferred from spectroscopy.  The unmagnified
inferred apparent magnitude of the lens sample members is then
\begin{equation}
  m_{inf} =  m_{obs} + 2.5 \log{\mu_{inf}}  =  m_{obs} + 1.0857\ln{\mu_{inf}} \quad .
\end {equation}
%ln  x = 2.30258 log x
%log x = 0.43429 ln  x
%2.5*log x = 1.08573 ln x

Two factors influence the ratio of the probability $P(\ln \mu)$ of true
logarithmic magnifcation $\ln \mu$ to the probability $P(\ln\mu_{inf})$
of inferred logarithmic magnification $\ln \mu_{inf}$.

The first of these reflects errors in determining $\ln \mu_{inf}$ from
the available data, which we  take to be a Gaussian with variance $\sigma_{\ln \mu}$.

The second of these arises from the number-magnitude relation,\footnote
{As we have assumed that the redshift of the source is known,
  the number-magnitude relation is entirely determined by the luminosity
  function, sidestepping the additional complications of angular diameter
  distances.}
which we take to be a power law in received flux, of the
form $10^{\beta (m - m_{inf})}$. 

Taken together, we have the probability of logarithmic magnification $P(\ln \mu)$ in
terms of  the probability of the inferred logarithmic magnification  $P(\ln \mu_{inf})$,
%\begin{align}  
%  \frac{P(\ln \mu)}{P(\ln \mu_{inf})} & = \Big[10^{1.0857 \beta(\ln\mu - \ln\mu_{inf})} \Big] \nonumber \\ 
%&\times \exp \left[-(\ln \mu -\ln \mu_{inf})^2 \over 2\sigma^2_{\ln\mu}\right]
%\,.  
%\end{align}

\begin{equation}
  \frac{P(\ln \mu)}{P(\ln \mu_{inf})}  = \Big[10^{1.0857 \beta(\ln\mu - \ln\mu_{inf})} \Big] \exp \left[-(\ln \mu -\ln \mu_{inf})^2 \over 2\sigma^2_{\ln\mu}\right]
\,.  
\end{equation}

The number-magnitude term has the effect of favoring larger values of
$\mu/\mu_{inf}$, in the same way the $A/A_{inf}$ term in
Section \ref{subsec:areabias}, favors larger caustic areas.  If magnification and
caustic area were uncorrelated, on might use this to compute an
``inferred magnification bias'' -- distinct from the
the ``magnification bias'' identified by \citet{turner} which
results from decreasing $\beta$ with fainter apparent magnitudes.

Instead we incorporate a factor for the caustic area bias that
we have heretofore suppressed and
proceed to compute a ``correlated area-magnification bias.''
%\begin{align}
%\frac{P(\ln \mu)}{P(\ln \mu_{inf})}& =\left[\frac{A}{A_{inf}}\right] \Big[10^{1.0857 \beta(\ln\mu - \ln\mu_{inf})} \Big] \nonumber \\
%&\times  \exp \left[-(\ln \mu -\ln \mu_{inf})^2 \over 2\sigma^2_{\ln\mu}\right]
%  \quad .
%\end{align}
\begin{equation}
\frac{P(\ln \mu)}{P(\ln \mu_{inf})} = \left[\frac{A}{A_{inf}}\right] \Big[10^{1.0857 \beta(\ln\mu - \ln\mu_{inf})} \Big] \exp \left[-(\ln \mu -\ln \mu_{inf})^2 \over 2\sigma^2_{\ln\mu}\right]
\quad .
\end{equation}
Adopting our assumption of perfect inverse correlation between magnification
and caustic area, we have
%\begin{align}
%\frac{P(\ln \mu)}{P(\ln \mu_{inf})}& = \left[\frac{\mu}{\mu_{inf}}\right]^{-\frac{1}{\alpha}}\Big[10^{1.0857 \beta(\ln\mu - \ln\mu_{inf})} \Big]  \nonumber \\
%&\times \exp \left[-(\ln \mu -\ln \mu_{inf})^2 \over 2\sigma^2_{\ln\mu}\right]
%  \quad .
%\end{align}
\begin{equation}
  %\frac{P(\ln \mu)}{P(\ln \mu_{inf})} = \left[\frac{\mu}{\mu_{inf}}\right]^{-\frac{1}{\alpha}}\Big[10^{1.0857 \beta(\ln\mu - \ln\mu_{inf})} \Big]
  \frac{P(\ln \mu)}{P(\ln \mu_{inf})} = \left[\frac{\mu}{\mu_{inf}}\right]^{-1/\alpha}\Big[10^{1.0857 \beta(\ln\mu - \ln\mu_{inf})} \Big]
%\frac{P(\ln \mu)}{P(\ln \mu_{inf})} = \left[\frac{\mu}{\mu_{inf}}\right]^{-1/\alpha}10^{\wedge}  \Big[{1.0857 \beta(\ln\mu - \ln\mu_{inf})} \Big]    
\exp \left[-(\ln \mu -\ln \mu_{inf})^2 \over 2\sigma^2_{\ln\mu}\right]
\quad .
\end{equation}
Incorporating all of these into a single exponential, we have
%\begin{align}
%\frac{P(\ln \mu)}{P(\ln \mu_{inf})} & = \nonumber \\
% \exp \Big\{\frac{1}{2\sigma^2_{\ln\mu}}\bigl[ & 
%    - \frac{2}{\alpha} \sigma^2_{\ln \mu}(\ln{\mu} - { \ln \mu_{inf}}) \nonumber \\      
%&  + 5 \beta \sigma^2_{\ln \mu} (\ln\mu - \ln\mu_{inf}) 
%  - (\ln \mu -\ln \mu_{inf})^2 \bigr]\Big\}\,.
%\end{align}
\begin{equation}
\frac{P(\ln \mu)}{P(\ln \mu_{inf})}  = 
 \exp \Big\{\frac{1}{2\sigma^2_{\ln\mu}}\bigl[  
     - \frac{2}{\alpha} \sigma^2_{\ln \mu}(\ln{\mu} - { \ln \mu_{inf}}) 
  + 5 \beta \sigma^2_{\ln \mu} (\ln\mu - \ln\mu_{inf}) 
  - (\ln \mu -\ln \mu_{inf})^2 \bigr]\Big\}\,.
\end{equation}

As in Section \ref{subsec:areabias} we can complete the square in the square
brackets, 
giving a  Gaussian whose peak value is shifted 
by the two power law terms.  
\begin{equation}\label{eqn:mucorrel}
\ln{\mu_{peak}} = \ln{\mu_{inf}} - \frac{1}{\alpha} \sigma^2_{\ln{\mu}} +2.5\beta\sigma^2_{\ln{\mu}}\quad . 
\end{equation}
In the (unrealistic) limiting case where the caustic area becomes
independent of the magnification, $\frac{1}{\alpha} \rightarrow 0$,
equation (\ref{eqn:mucorrel}) reduces to the
usual Malmquist correction, but with uncertainties in $\ln \mu$
instead of in magnitudes.

Alternatively, starting with Gaussian errors in caustic area, we have
\begin{equation}\label{eqn:acorrel}
   \ln A_{peak} =  \ln A_{inf} - \sigma^2_{\ln A} +2.5        \beta  \alpha \sigma^2_{\ln A}\quad ,   
\end{equation}
which would agree with equation (\ref{eqn:areapeak}) in the limit of $\beta =0 $, a flat number-magnitude relation.
    
Equations (\ref{eqn:mucorrel}) and (\ref{eqn:acorrel}) give a sense of
how inferred caustic area bias and inferred magnification bias
compensate for each other for the case of perfect anti-correlation.
As the parameter $\alpha$ depends upon the adopted lens model and
the parameter $\beta$ depends on details of the sample selection,
the extent of that compensation must be determined on a case by
case basis.


\newpage
\bibliographystyle{aasjournal}
\bibliography{Paper}



\end{document}